# Bright muon source driven by GeV electron beams from a compact laser wakefield accelerator

Bobbili Sanyasi Rao[1,#], Jong Ho Jeon[1], Hyung Taek Kim[1,2†], and Chang Hee Nam[1,3]

[1]Center for Relativistic Laser Science (CoReLS), Institute for Basic Science, Korea

[2]Advanced Photonics Research Institute, Gwangju Institute of Science and Technology (GIST), Korea

[3]Department of Physics and Photon Science, GIST, Korea

[#]sunnyb@rrcat.gov.in / [†]htkim@gist.ac.kr

We report here a systematic quantitative study on generation and characteristics of an active muon source driven by the interaction of an electron beam within the energy range of 1 – 10 GeV from laser wakefield acceleration (LWFA) with a tungsten target, using Monte Carlo simulations. The 10-GeV electron beam, achievable in near future, from LWFA using femtosecond multi-PW lasers is employed to drive the bright source of muon pairs in a compact setup. We show that a highly directional and intense source of short-pulsed GeV muon pairs ($\mu^-\mu^+$) having peak brightness $5 \times 10^{17}$ pairs/s/cm$^2$/sr and sub-100-ps duration could be produced using a quasi-monoenergetic 10-fs, 10-GeV electron bunch with 1-mrad divergence and 100-pC charge. The muon pairs are emitted from a point-like source with well-defined position and timing, and the source has size and geometric emittance about 1 mm and 40 µm, respectively. Such muon sources can greatly benefit applications in muon radiography, studies on anomalous dipole moment and rare decays of muons, neutrino oscillations, and an injector of a future compact muon collider.

## INTRODUCTION

Laser wakefield acceleration (LWFA) with ultra-high acceleration gradient ≥ 100 GV/m is a very attractive alternative to conventional rf-cavity based acceleration [1,2]. The LWFA has the potential to reduce the size and the cost of the future TeV electron-positron (e$^-$e$^+$) collider significantly [3]. There has been enormous progress in the field of LWFA, particularly in the last two decades by virtue of continuous advancement of ultrashort high-power laser technologies. Particularly, availability of femtosecond PW-class lasers in recent years has enabled the generation of multi-GeV electron beams from cm-scale plasma [4-8]. By the ongoing efforts with multi-PW lasers, the first 10-GeV electron beams from LWFA will be inevitably demonstrated soon. [9- 12]. Realization of stable 10 GeV LWFA has the potential to open up a plethora of new possibilities which include the generation of a bright source of muon pairs ($\mu^-\mu^+$) in a compact all optical setup [13]. There is a world-wide interest to develop intense sources of muons which can be applied to muon radiography, precision muon physics, muon-catalyzed fusion, neutrino oscillations and compact muon colliders [14-18]. Currently, muons sources are mainly driven by high energy proton accelerators [14, 17] which are much larger and complex in comparison to electron accelerators. There is a growing interest in recent years to develop compact muons sources based on electron accelerators [15]. The LWFA is a quite attractive alternative to conventional accelerators as they promise significant reduction in the size, which might help establish muon facilities even in small-scale laboratories in future. Furthermore, multi-GeV LWFA may enable direct generation of muon pairs [16] with unique properties of high directionality, sub-100ps duration, and excellent emittance from an all optical table-top setup. The laser-driven muon sources are also scalable with laser power and may have synergies with nuclear fusion [17].

The investigation on the feasibility of a compact muon source based on a high energy laser electron accelerator is an important task as the first step for the development of the source. The possibility of muon pair-production driven by GeV electron beams from LWFA has been first considered by A. I. Titov et al. [13]. A detailed quantitative analysis of the muon pair production and the source characteristics under realistic interaction conditions, however, has not been done yet. Here, we employ particle tracking simulations based on a Monte Carlo method using G4beamline [19, 20] and Geant4 toolkits [21,22] that consider all the important physical processes relevant to the incident electron beam energy and produce detailed practical results from the interaction. Using the simulations, we present here a systematic quantitative study of an active source of muon pairs driven by the interaction of a quasi-monoenergetic electron beam with a tungsten target, in the incident electron energy range from 1 to 10 GeV with typical beam parameters of LWFA. We showed that the yield of muon pairs increases almost linearly with electron energy in the energy range from 1 to 10 GeV. We also present the characteristics of the muon pairs generated for different electron beam energies, demonstrating the feasibility to produce a bright, very short pulsed source of GeV muon pairs with peak brightness of about $5 \times 10^{17}$ pairs/s.cm$^2$.sr and duration 80 ps using a 10fs, 10 GeV electron bunch with a beam charge of 100 pC interacting with a 10 mm thick tungsten target. The muon source is well localized within ≲1 mm, well synchronized with respect to the drive laser, and exhibits low emittance ≈ 40 μm. The quantitative estimates and analysis from our study will be helpful to design experiments for producing an intense source of muons, to develop appropriate detection methods, and also to employ the source for applications.

**SETUP FOR MUON PAIR PRODUCTION AND SIMULATIONS**

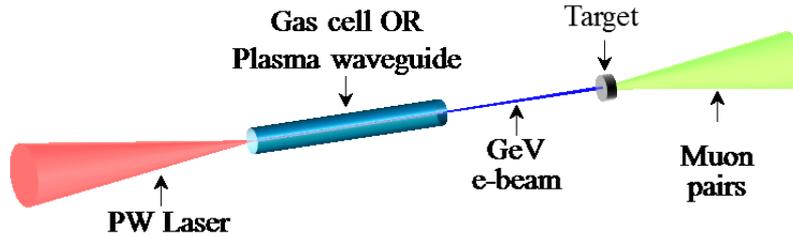

Fig. 1 Schematics for producing highly directional GeV muon pairs using a laser wakefield accelerator

We investigated the muon pair creation based on a realistic physical scenario with current or upcoming laser electron accelerators. A schematic setup for production of muon pairs ($\mu^-\mu^+$) driven by GeV-LWFA is shown in Fig.1. A PW-class fs laser pulse drives wakefield acceleration of electrons from a several-cm scale gas cell medium or pre-formed plasma wave-guide and generates a GeV electron beam. The electron beam interaction with a high-Z target produces muon pairs in the forward direction. The pair production takes place in two-step process: firstly the electrons ($e^-$) interact with the target nuclei (N) and produce a broad spectrum of photons (γ) with energy ($E_\gamma$) extending up to the incident electron energy by bremsstrahlung process, and secondly the interaction of photons with the target nuclei produces muon pairs, as represented below, when the photon energy, $E_\gamma > 2m_\mu c^2 \approx 211$ MeV (where $m_\mu$ is the rest mass energy of muon).

$$\left.\begin{array}{r} e^- + N \rightarrow e^{-*} + N + \gamma \\ \gamma + N \rightarrow N + \mu^-\mu^+ \end{array}\right\} \quad (1)$$

The second process is the principal mechanism responsible for the muon pair production here, which is analogous to the well-known Bethe-Heitler process of electron-positron pair production [23].

We performed Monte Carlo simulations using G4beamline (version 3.02.1) and Geant4 physics toolkits to quantitatively analyze the generation and characterization of the muon source driven by a GeV electron beam interacting with a solid target. The physics list QGSP_BERT [24, 25] has been used and the gamma to muon pair conversion (in the presence of nuclei) was activated for the simulations. Since the threshold for muon pair production is 211 MeV, it is necessary to have an electron beam with several times the threshold energy with a reasonably achievable charge (~100 pC) to be able to generate a significant flux of high energy bremsstrahlung photons which will further produce a sizeable number of muon pairs. We consider here an electron beam with peak energy in the range from 1 to 10 GeV for simulating muon pair production and to study their characteristics. We define electron beam at the exit of LWFA (e$^-$-source) with the Gaussian distribution having root-mean-square (rms) values of transverse size 10 μm, bunch length 10 fs, divergence 1 mrad, and relative energy spread 5%. These features may be considered typical to quasi-monoenergetic electron beams from LWFA. Tungsten (W), due to its high atomic number, Z, is chosen as a target for muon pair production, which enables efficient production of bremsstrahlung photon for high yield of muon pairs consequently [23]. The geometry of the target is considered to be disc-shaped with diameter 2.5 cm and variable thickness from 0.1 to 1.8 cm (radiation length, $X_0$ of tungsten is ≈ 3.3 mm). The target was placed in the electron beam path, as shown in the Fig. 1, at a distance of 50cm from the e$^-$-source. To identify and measure different parameters of the various particles and photons generated from the interaction we used flat surfaced virtual detectors at 3 locations, viz. (1) at 0.4 m from the e$^-$-source before the beam strikes the Tungsten target, (2) at 0.51 m from the source, immediately after the target (5 mm from the target exit surface), and (3) at 1.5 m from the e$^-$-source or 1m from the Tungsten target. We analyzed the characteristics of the muon beam from the target as well as incident electron beam before and after the interaction with the Tungsten target as described in following sections.

**YIELD OF MUON PAIRS**

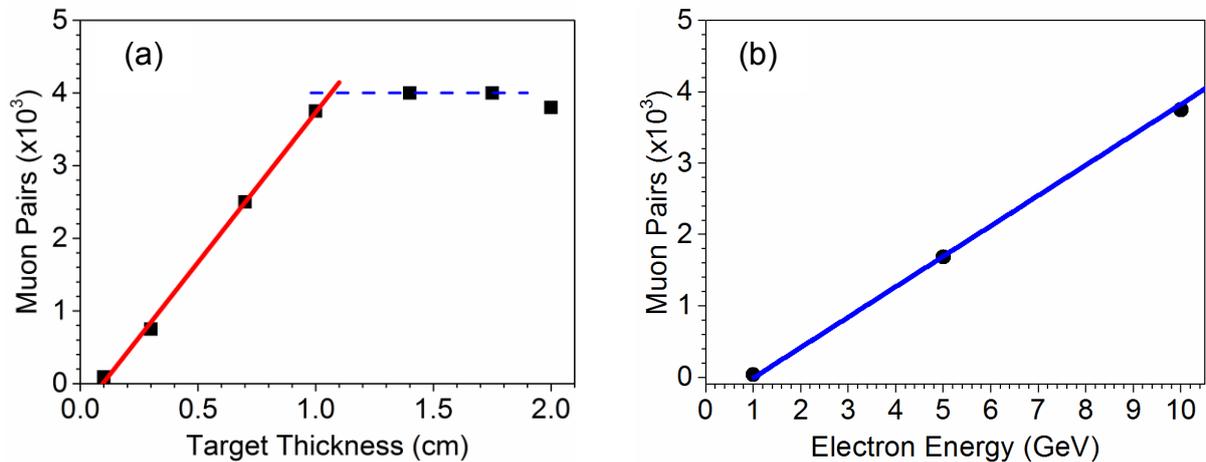

Fig. 2 Dependence of the yield of muon pairs (a) on target thickness for the incident electron beam of 10 GeV energy and (b) on the incident electron beam energy for 1cm thick tungsten target. The muon yield shown here is for the incident electron beam charge of 100 pC.

The muon pair production was simulated for a 10 GeV electron beam interacting with W-target of different thicknesses. The yield of muon pairs for the incident electron beam charge of 100 pC obtained as a function of the target thickness is shown in Fig. 2(a). The yield increases from close to 100 pairs for 1mm to about 3800 pairs for 10 mm (≈3$X_0$). Further increase in the thickness shows saturation at 4000 pairs as indicated by a dashed line in blue. The yield from 1mm to 10mm thickness fits well to linear scaling, as shown in the red line of Fig. 2, with a growth rate ~ 410 pairs/mm. By the definition of radiation length, it is expected that the electron beam loses almost all of its initial energy within the target thickness of $3X_0$, which could cause the saturation. This suggests that the thickness beyond the saturation length cannot increase the yield, while it can lead to increase the size and degrade the source quality due to straggling effects. It can be noted from the Fig. 2(a) that increasing the thickness well beyond $3X_0$ can also reduce the total yield due to absorption of low energy muon pairs in the target. Particularly at a low energy of the incident electron beam both straggling and absorption could seriously degrade the muon source. Therefore, the target thickness could be optimized at about 3 times of the radiation length to generate a maximum number of muon pairs without sacrificing the beam quality.

By fixing the W-target thickness to its optimum value near 1cm ($\approx 3X_0$), we simulated the muon pair production for the incident electron beam with different peak energy in the range from 1 to 10 GeV. Figure 2(b) shows the yield of muon pairs as a function of the incident electron energy. The yield was about 40 pairs at 1 GeV and increased linearly with respect to the energy at a growth rate of ~ 425 pairs/1GeV as shown in the Fig. 2(b). The linear dependence on electron energy suggests that the desired yield of muon pairs could be achieved either by increasing the electron beam energy at a constant charge or by increasing the beam charge at a constant beam energy. However, the generated muon pairs have smaller emission angle and higher kinetic energy for higher incident electron beam energy, compared to those produced using low energy electron beam, as elucidated in the next section.

## CHARACTERISTICS OF THE MUON SOURCE

The knowledge on the emission characteristics of a photon or particle source is necessary for its utility in any application. Figures 3(a) and (b) show the angular distribution of all the muon pairs ($\mu^-\mu^+$) produced when an electron beam with different energies (2.5, 5, and 10 GeV) and constant charge strikes 1 cm thick target. It is quite clear that the angular distribution becomes narrower along the forward direction with the increase of incident electron energy. The angular distribution fits quite well to a Lorentzian-peaked distribution. Figures 3(c) and (d) show angle-integrated energy spectra of all the muon pairs produced from the target. The spectra have peaked distribution with a gently falling long tail extending close to the incident electron beam energy. The peaks are centered near 0.5 GeV for the incident electron beam energies 2.5, 5, and 10 GeV. The muon spectrum for 10 GeV electron beam contains a significant fraction of energy in excess of 2 GeV, while the bulk of the muons pairs are concentrated within 2 GeV energy for the 3 incident beam energies. This result shows that we can obtain highly energetic and angularly confined muon beam from the interaction between a multi-GeV electron beam and a high Z material.

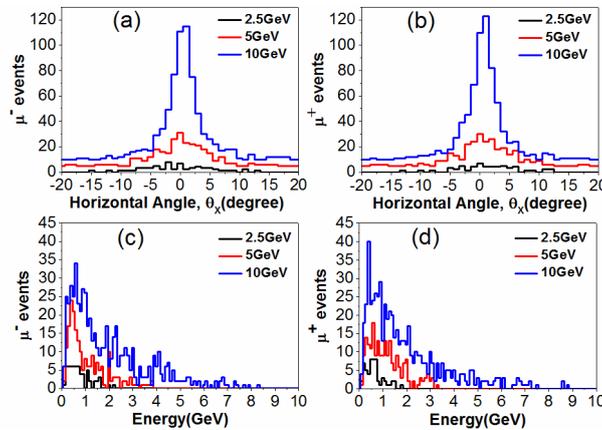

Fig. 3 Angular distribution of (a) $\mu^-$ and (b) $\mu^+$ and energy distribution of (c) $\mu^-$ and (d) $\mu^+$ produced from muon pair production in 1cm thick tungsten target when an electron beam with different peak energies (2.5, 5, and 10 GeV) and constant charge of $1\times10^8$ e$^-$ interacts with the target. For better clarity, the base-lines of the angular distribution of the muons are shifted by $5^0$ for 5 GeV and $10^0$ for 10 GeV w.r.t. the x-axis.

The muon pairs produced from the Bethe-Heitler process exhibit characteristic angular spread [23] of the order of $1/\gamma_\mu$, where $\gamma_\mu = E_\mu/m_\mu c^2$ is the relativistic gamma factor corresponding to the maximum energy of the muons, $E_\mu \approx E_e$ (peak energy of the incident electron beam). The full width at half maximum (FWHM) angular spread of the muon pairs is obtained from a Lorentzian fit to the angular distribution and its variation with electron energy is shown in Fig. 4. In Fig. 4, we also show curves representing $4/\gamma_\mu$ (solid line) and $6/\gamma_\mu$ (dashed line) values for different incident electron energy. The angular spread fits quite well with the $6/\gamma_\mu$ curve for high electron energy ($\gtrsim 5$ GeV) and deviates considerably from this for low electron energy. In particular, the muon pairs generated with 2.5GeV electron beam showed angular spread close to $4/\gamma_\mu$. This suggests that while a simple linear dependence of angular spread on $1/\gamma_\mu$ works well for electron energy greater than >5GeV, where the muon pair production becomes practically more interesting, such a simple scaling might not work for the electron energy much lower than 5GeV due to considerable deviation in the dynamics of muon pair production and strong straggling effect.

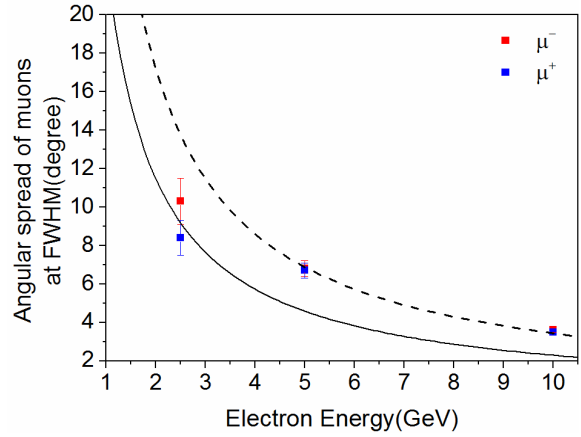

Fig. 4 Angular spread of the muon pairs produced from 1 cm thick W-target for different incident electron beam energies.

We investigated the energy-dependent angular distribution of the muon pairs for the 3 incident energies of the electron beam. As shown in Fig. 5, it is very clear that high energy muons are emitted within a narrow angle while low energy muons are emitted over a broader range of angles. The muons emitted along the incident electron beam axis have a broad spectrum of muons ranging from about 0 to maximum value close to the incident electron energy. However, the muons emitted off-axis have a narrower range and reduced maximum energy. This fact could be exploited to design systems for detection of the muons pairs and also for filtering the muon pairs either using magnetic optics or employing apertures according to intended applications.

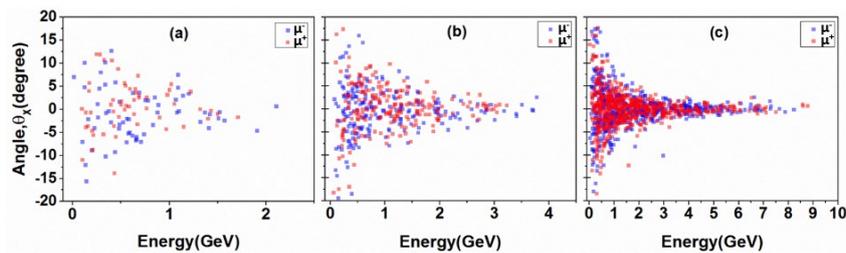

Fig. 5 Energy-dependent angular distribution of muon pairs ($\mu^-\mu^+$) produced when an electron beam with energy (a) 2.5 GeV, (b) 5 GeV, and (c) 10 GeV and charge of $1\times10^8$ $e^-$ interacts with a 1 cm thick W-target.

Since the high energy electron beam tends to produce not only high yield of muon pairs but also a high degree of directionality, a further analysis of the temporal and phase space distribution of the muons is performed for the case of 10 GeV incident electron beam. Figure 6(a) shows the temporal distribution of the muon pairs measured at 1.5 m from the source. The number of muon pairs sharply rises at 5 ns, the transit time for the relativistic particles to reach the detector from the source, and followed by relatively slow fall. As shown by the black dashed line in Fig. 6(a), the decay could be approximated by an exponential fit with 1/e falling time, $\Delta t$ = 26 ps, and most of generated muon pairs (> 90%) are concentrated within 80 ps ($\approx 3\Delta t$). Figure 6(b) shows the phase space distribution of the muon pairs measured at 5 mm from the exit surface of the target. From the phase space coordinates of all the particles produced we estimate the rms source size of 1mm and the geometrical emittance ~ 40 μm. The source with such low emittance may be well suited for high-resolution muon radiography. Considering the maximum yield of 4000 muon pairs from a 1mm source within 80 ps and $3.6^0$ emission angle the peak brightness of the source is estimated to be about $5 \times 10^{17}$ pairs/s.cm$^2$.sr when a 10 GeV electron beam interacts with 1cm thick tungsten. It may be noted that the muon beam emittance could be lower and the brightness could be higher in the actual case with 10 GeV electron beam from LWFA than those presented here, as it may have lower divergence than considered here.

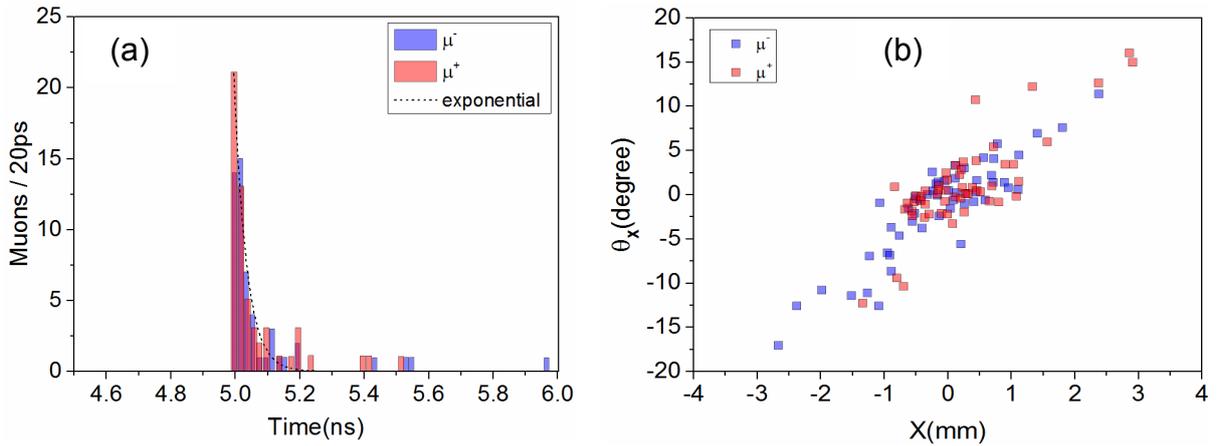

Fig. 6 Temporal (a) and phase-space (b) distribution of muon-pairs generated when an electron beam with energy 10 GeV and charge $1 \times 10^7 e^-$ incident on 1 cm thick W-target. The temporal distribution was measured at 1.5 m from the exit of LWFA (source of the electron beam). The phase space coordinates were measured near the exit surface of the W target.

**DISCUSSION**

The possibility of producing leptons, hadrons, and neutrinos driven by energetic electrons /ions from ultra-intense laser plasma interactions has been suggested earlier [26, 27]. Particularly, the possibility of muon pair production using the electron beams driven by LWFA has been first reported with a qualitative analysis of the muon source and an order of magnitude estimates of muon yield [13]. However, these estimates consider that the bremsstrahlung photons contribute entirely to the production of muon pairs, and ignore the electron-positron pair creations which are by far the dominant channel of photon loss mechanism in the interaction, apart from other processes such as the production of protons, neutrons, and pions. Such an assumption might lead to overestimation of a total number of muon pairs compared to the realistic scenario as observed in our simulations. For comparison, the predicted muon yield in Ref. [13] is about 5 times higher than that observed from our simulations. Here, the simulations have been conducted by taking into consideration of all the major channels of bremsstrahlung photon depletion along with absorption and straggling of muons which can also affect the yield and the quality of the source, particularly at incident electron energy close to 1 GeV. There has been some interest to explore the possibility of developing a compact muon source using electron-photon collisions using LWFA [28]. However, such collisions produce a significantly lesser yield of muons for a given driving laser energy compared to electron-target collisions and they require also inherently complex implementation owing to stringent requirements of spatial and temporal overlaps of electron-photon beams. Our scheme, the bombardment of a high energy electron beam on a high Z material, can be a simple and robust method to obtain a bright muon source for various applications.

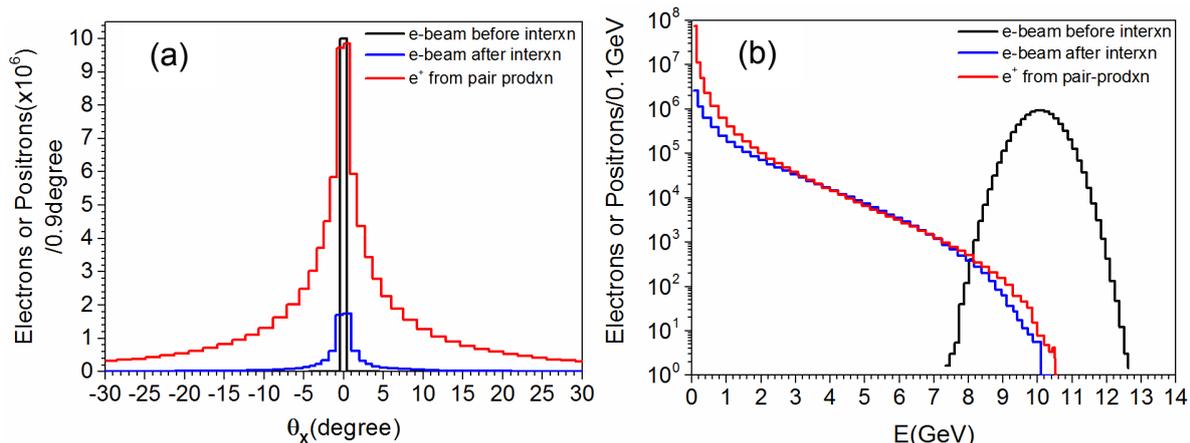

Fig. 7 Angular (a) and energy (b) distribution of the electrons and positrons in the forward direction when a 10±0.5GeV (1.6 pC) electron beam with 1mrad divergence incident on 1cm thick tungsten target.

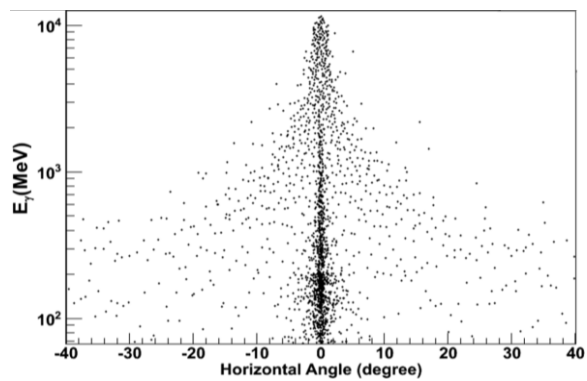

Fig. 8 Energy-dependent angular spread of photons emitted in the forward direction when an electron beam with energy of 10±0.5GeV, charge of 1.6 pC, and 1mrad divergence is incident on 1cm thick tungsten target.

The simulation results presented here consider real photons produced by the bremsstrahlung process. Actually, virtual photons associated with the incident electron beam can also produce particle pairs much like real photons. The yield from this process is, however, negligible compared to that by the real photons when target thickness $\gg X_0/25$ ($X_0$ is the radiation length of the target) [13, 23], which is the case of our study. In addition, the yield of muon pairs at electron energy close to the threshold energy is negligible because the cross-section is small and produced pairs cannot travel through the target. Further, the small energy of the produced muon pairs for incident electron energy close to threshold makes the saturation of the yield earlier than 3 $X_0$. Although the incident electron beam with energy in excess of 1 GeV can produce a sizeable number of muon pairs, the source produced with electrons with energy near and above 10 GeV is practically more interesting due to the high degree of collimation of the muon beam. Therefore, a 10-GeV electron beam from LWFA can be a good candidate as a driver for a high-quality source of both negative and positive muons.

When a high energy electron beam interacts with a high-Z target, the tertiary particles like muon pairs are generated in the background of a significant number of photons and particles. The most dominant background in the direction of muon emission arises due to electrons, positrons, and bremsstrahlung photons from the target. We have shown in Fig. 7 the angular and energy distributions of electron-positron pairs produced from the interaction of an incident electron beam with a 1cm thick tungsten target. The electrons from the pair production exhibited similar properties to positrons. Figure 8 shows the angular-dependent energy distribution of the bremsstrahlung photons emitted from the interaction. The number and energy of the incident electron beam are attenuated significantly in the interaction with 1 cm thick target while producing a significant flux of electron-positron pairs and photons with energy of a few 100 MeV in a narrow-angle. One way to detect the muons is the selection of an appropriate combination of low-Z and high-Z materials ($\gtrsim 3\ X_0$ thickness) to effectively diminish the background of electrons, positrons, photons, and neutrons (other particles with relatively small population, like p,$\pi^{\pm,0}$, can also be easily taken care by such shielding), but allows the muon pairs for detection using a combination of plastic scintillator and high sensitivity intensified CCD camera with gating [16]. Another method could be to disperse the electrons and positrons from the incident axis by a magnetic field. This may significantly reduce the bremsstrahlung photons in the beam path when they interact with the low-Z/high-Z combination materials for filtering, and allow the muon pair with minimum deflection (due to the much higher relative mass of muon pairs compared to $e^-/e^+$) to reach the detector. Consequently, we should carefully design the muon detection system such that it should minimize the effects of background emissions from energetic electrons and photons.

**SUMMARY**

We have carried out a detailed study of the muon pair production using Monte Carlo simulations for quantitative analysis and understanding of various secondary, tertiary particles and high energy photons from the interaction of a GeV electron beam with a solid tungsten target. Particular emphasis was laid on the optimization of muon/anti-muon pair creations and their characterization. We note that a bright source of short-pulsed muon beam with 100s of MeV up to GeV energy could be produced using multi-GeV electron beams. The multi-GeV to 10 GeV electron beams produced with multi-PW lasers in future can be applied to drive intense beam of muon pairs. The simulations also revealed that a significant background of high energy photons and electron-positron pairs, in addition to exotic particles like pions, can be produced from the interaction. The present study will be greatly beneficial for designing experiments for the optimal generation and unambiguous detection of muon pairs driven by electron beams with energy of multi-GeV to near 10 GeV in near future.

**Acknowledgment**

This work was supported by the Institute for Basic Science (IBS-R012-D1) and the Research on Advanced Optical Science and Technology grant funded by GIST.